

\font\bigbold=cmbx12
\font\ninerm=cmr9      \font\eightrm=cmr8    \font\sixrm=cmr6
\font\fiverm=cmr5
\font\ninebf=cmbx9     \font\eightbf=cmbx8   \font\sixbf=cmbx6
\font\fivebf=cmbx5
\font\ninei=cmmi9      \skewchar\ninei='177  \font\eighti=cmmi8
\skewchar\eighti='177  \font\sixi=cmmi6      \skewchar\sixi='177
\font\fivei=cmmi5
\font\ninesy=cmsy9     \skewchar\ninesy='60  \font\eightsy=cmsy8
\skewchar\eightsy='60  \font\sixsy=cmsy6     \skewchar\sixsy='60
\font\fivesy=cmsy5     \font\nineit=cmti9    \font\eightit=cmti8
\font\ninesl=cmsl9     \font\eightsl=cmsl8
\font\ninett=cmtt9     \font\eighttt=cmtt8
\font\tenfrak=eufm10   \font\ninefrak=eufm9  \font\eightfrak=eufm8
\font\sevenfrak=eufm7  \font\fivefrak=eufm5
\font\tenbb=msbm10     \font\ninebb=msbm9    \font\eightbb=msbm8
\font\sevenbb=msbm7    \font\fivebb=msbm5
\font\tenssf=cmss10    \font\ninessf=cmss9   \font\eightssf=cmss8
\font\tensmc=cmcsc10

\newfam\bbfam   \textfont\bbfam=\tenbb \scriptfont\bbfam=\sevenbb
\scriptscriptfont\bbfam=\fivebb  \def\Bbb{\fam\bbfam}
\newfam\frakfam  \textfont\frakfam=\tenfrak \scriptfont\frakfam=%
\sevenfrak \scriptscriptfont\frakfam=\fivefrak  \def\frak{\fam\frakfam}
\newfam\ssffam  \textfont\ssffam=\tenssf \scriptfont\ssffam=\ninessf
\scriptscriptfont\ssffam=\eightssf  
\def\smc{\tensmc}

\def\eightpoint{\textfont0=\eightrm \scriptfont0=\sixrm
\scriptscriptfont0=\fiverm  \def\rm{\fam0\eightrm}%
\textfont1=\eighti \scriptfont1=\sixi \scriptscriptfont1=\fivei
\def\oldstyle{\fam1\eighti}\textfont2=\eightsy
\scriptfont2=\sixsy \scriptscriptfont2=\fivesy
\textfont\itfam=\eightit         \def\it{\fam\itfam\eightit}%
\textfont\slfam=\eightsl         \def\sl{\fam\slfam\eightsl}%
\textfont\ttfam=\eighttt         \def\tt{\fam\ttfam\eighttt}%
\textfont\frakfam=\eightfrak     \def\frak{\fam\frakfam\eightfrak}%
\textfont\bbfam=\eightbb         \def\Bbb{\fam\bbfam\eightbb}%
\textfont\bffam=\eightbf         \scriptfont\bffam=\sixbf
\scriptscriptfont\bffam=\fivebf  \def\bf{\fam\bffam\eightbf}%
\abovedisplayskip=9pt plus 2pt minus 6pt   \belowdisplayskip=%
\abovedisplayskip  \abovedisplayshortskip=0pt plus 2pt
\belowdisplayshortskip=5pt plus2pt minus 3pt  \smallskipamount=%
2pt plus 1pt minus 1pt  \medskipamount=4pt plus 2pt minus 2pt
\bigskipamount=9pt plus4pt minus 4pt  \setbox\strutbox=%
\hbox{\vrule height 7pt depth 2pt width 0pt}%
\normalbaselineskip=9pt \normalbaselines \rm}

\def\ninepoint{\textfont0=\ninerm \scriptfont0=\sixrm
\scriptscriptfont0=\fiverm  \def\rm{\fam0\ninerm}\textfont1=\ninei
\scriptfont1=\sixi \scriptscriptfont1=\fivei \def\oldstyle%
{\fam1\ninei}\textfont2=\ninesy \scriptfont2=\sixsy
\scriptscriptfont2=\fivesy
\textfont\itfam=\nineit          \def\it{\fam\itfam\nineit}%
\textfont\slfam=\ninesl          \def\sl{\fam\slfam\ninesl}%
\textfont\ttfam=\ninett          \def\tt{\fam\ttfam\ninett}%
\textfont\frakfam=\ninefrak      \def\frak{\fam\frakfam\ninefrak}%
\textfont\bbfam=\ninebb          \def\Bbb{\fam\bbfam\ninebb}%
\textfont\bffam=\ninebf          \scriptfont\bffam=\sixbf
\scriptscriptfont\bffam=\fivebf  \def\bf{\fam\bffam\ninebf}%
\abovedisplayskip=10pt plus 2pt minus 6pt \belowdisplayskip=%
\abovedisplayskip  \abovedisplayshortskip=0pt plus 2pt
\belowdisplayshortskip=5pt plus2pt minus 3pt  \smallskipamount=%
2pt plus 1pt minus 1pt  \medskipamount=4pt plus 2pt minus 2pt
\bigskipamount=10pt plus4pt minus 4pt  \setbox\strutbox=%
\hbox{\vrule height 7pt depth 2pt width 0pt}%
\normalbaselineskip=10pt \normalbaselines \rm}

\global\newcount\secno \global\secno=0 \global\newcount\meqno
\global\meqno=1 \global\newcount\appno \global\appno=0
\newwrite\eqmac \def\romappno{\ifcase\appno\or A\or B\or C\or D\or
E\or F\or G\or H\or I\or J\or K\or L\or M\or N\or O\or P\or Q\or
R\or S\or T\or U\or V\or W\or X\or Y\or Z\fi}
\def\eqn#1{ \ifnum\secno>0 \eqno(\the\secno.\the\meqno)
\xdef#1{\the\secno.\the\meqno} \else\ifnum\appno>0
\eqno({\rm\romappno}.\the\meqno)\xdef#1{{\rm\romappno}.\the\meqno}
\else \eqno(\the\meqno)\xdef#1{\the\meqno} \fi \fi
\global\advance\meqno by1 }

\global\newcount\refno \global\refno=1 \newwrite\reffile
\newwrite\refmac \newlinechar=`\^^J \def\ref#1#2%
{\the\refno\nref#1{#2}} \def\nref#1#2{\xdef#1{\the\refno}
\ifnum\refno=1\immediate\openout\reffile=refs.tmp\fi
\immediate\write\reffile{\noexpand\item{[\noexpand#1]\ }#2\noexpand%
\nobreak.} \immediate\write\refmac{\def\noexpand#1{\the\refno}}
\global\advance\refno by1} \def\semi{;\hfil\noexpand\break ^^J}
\def\nl{\hfil\noexpand\break ^^J} \def\refn#1#2{\nref#1{#2}}

\def\cmp#1#2#3{{\it Commun.\ Math.\ Phys.}\ {\bf {#1}} ({#2}) #3}

\def\plA#1#2#3{{\it Phys.\ Lett.}\ {\bf {#1}A} ({#2}) #3}

\def\prl#1#2#3{{\it Phys.\ Rev.\ Lett.}\ {\bf #1} ({#2}) #3}

\newif\iftitlepage \titlepagetrue \newtoks\titlepagefoot
\titlepagefoot={\hfil} \newtoks\otherpagesfoot \otherpagesfoot=%
{\hfil\tenrm\folio\hfil} \footline={\iftitlepage\the\titlepagefoot%
\global\titlepagefalse \else\the\otherpagesfoot\fi}

\def\abstract#1{{\parindent=30pt\narrower\noindent\ninepoint\openup
2pt #1\par}}

\newcount\notenumber\notenumber=1 \def\note#1
{\unskip\footnote{$^{\the\notenumber}$} {\eightpoint\openup 1pt #1}
\global\advance\notenumber by 1}

\def\today{\ifcase\month\or January\or February\or March\or
April\or May\or June\or July\or August\or September\or October\or
November\or December\fi \space\number\day, \number\year}

\def\pagewidth#1{\hsize= #1}  \def\pageheight#1{\vsize= #1}
\def\hcorrection#1{\advance\hoffset by #1}
\def\vcorrection#1{\advance\voffset by #1}

\font\extra=cmss10 scaled \magstep0  \setbox1 = \hbox{{{\extra R}}}
\setbox2 = \hbox{{{\extra I}}}       \setbox3 = \hbox{{{\extra C}}}
\setbox4 = \hbox{{{\extra Z}}}       \setbox5 = \hbox{{{\extra N}}}

\def\CCC{{{\extra C}}\hskip-\wd3\hskip 2.5 true pt{{\extra I}}
\hskip-\wd2\hskip-2.5 true pt\hskip\wd3}
\def\Complex{\hbox{{\extra\CCC}}\!\!}   
\def\ZZZ{{{\extra Z}}\hskip-\wd4\hskip 2.5 true pt{{\extra Z}}}
\def\Zed{\hbox{{\extra\ZZZ}}}           


\def\frac#1#2{{#1\over#2}}

\def\pmb#1{\setbox0=\hbox{$#1$} \kern-.025em\copy0\kern-\wd0
    \kern.05em\copy0\kern-\wd0 \kern-.025em\raise.0433em\box0 }

\def\ve{\vfill\eject}

\def\Z{{\Zed}}
\def\R{{\Real}\!}
\def\C{{\Complex}}
\def\({\left(}
\def\){\right)}


\def\ddef#1#2#3#4{ \left\{ \matrix{ #1 & #2 \cr #3 & #4 } \right. }

\def\d{{\rm d}}

\def\qr{^{(+)}}

\def\==>{\Longrightarrow}
\def\[{\left[}
\def\]{\right]}

\def\R{{\rm I \hskip-0.47ex R}}
\def\bit{\hskip 0.15ex}    
\def\bitt{\hskip 0.30ex}   
\def\bittt{\hskip 0.45ex}  
\def\bitttt{\hskip 0.60ex} 
\def\qD{{\cal D}}
\def\qF{{\cal F}}
\def\qH{{\cal H}}
\def\qP{{\cal P}}
\def\qX{{\cal X}}
\def\qidH{{\rm id}_\qH}
\def\ie{{\it i.e.,\ }}


{

\refn\AG
{N.I. Akhiezer and I.M. Glazman,
\lq\lq Theory of Linear Operators in Hilbert Space\rq\rq,
{\sl Vol.II},
Pitman Advanced Publishing Program, Boston, 1981}

\refn\ABD
{S. Albeverio, Z. Brze\'{z}niak and L. Dabrowski,
{\it J. Funct. Anal.} {\bf 130} (1995) 220}

\refn\ADK
{S. Albeverio, L. Dabrowski and P. Kurasov,
{\it Lett. Math. Phys.} {\bf 45} (1998) 33}

\refn\AGHH
{S. Albeverio, F. Gesztesy, R. H{\o}egh-Krohn and H. Holden,
\lq\lq Solvable Models in Quantum Mechanics\rq\rq,
Springer, New York, 1988}

\refn\AK
{S. Albeverio and P. Kurasov,
\lq\lq
Singular Perturbations of Differential Operators\rq\rq,
Cambridge Univ. Press,
Cambridge, 2000}

\refn\CHa
{T. Cheon,
\plA{248}{1998}{285}}

\refn\CFT
{T. Cheon, T. F\"{u}l\"{o}p and I. Tsutsui,
{\it
Symmetry, Duality and Anholonomy
of Point Interactions in One Dimension},
KEK Preprint 2000-54, quant-ph/0008123}

\refn\CSb
{T. Cheon and T. Shigehara,
\prl{82}{1999}{2536}}

\refn\ER
{C. Emmrich and H. R{\"o}mer,
\cmp{129}{1990}{69}}

\refn\EG
{P. Exner and H. Grosse,
{\it Some properties of the one-dimensional
generalized point
interactions (a torso)},
math-ph/9910029}

\refn\FT
{T. F\"{u}l\"{o}p and I. Tsutsui,
\plA{264}{2000}{366}}

\refn\Junker
{G. Junker,
\lq\lq Supersymmetric Methods in Quantum
and Statistical
Physics\rq\rq,
Springer,
Berlin, 1996}

\refn\Seba
{P. \v{S}eba,
{\it Czech. J. Phys.} {\bf 36} (1986) 667}

\refn\TFC
{I. Tsutsui, T. F\"{u}l\"{o}p and T. Cheon,
 {\it J. Phys. Soc. Jpn.} {\bf 69} (2000) 3473}

}



\pageheight{23cm}
\pagewidth{14.8cm}
\hcorrection{0mm}
\magnification= \magstep1
\def\bsk{%
\baselineskip= 16.8pt plus 1pt minus 1pt}
\parskip=5pt plus 1pt minus 1pt
\tolerance 6000


\null

{
\leftskip=100mm
\hfill\break
KEK Preprint 2001-12
\hfill\break
\par
}

\vskip 15pt

{\baselineskip=18pt

\centerline{\bigbold
M{\" o}bius Structure of the Spectral Space}
\centerline{\bigbold
of Schr{\"o}dinger Operators with Point Interaction}

\vskip 20pt

\centerline{\smc
Izumi Tsutsui\footnote{${}^*$}
{\eightpoint email:\quad izumi.tsutsui@kek.jp}
}

\vskip 5pt

{
\baselineskip=13pt
\centerline{\it
Institute of Particle and Nuclear Studies}
\centerline{\it
High Energy Accelerator Research Organization (KEK)}
\centerline{\it
Tsukuba 305-0801, Japan}
}

\vskip 3pt

\centerline{\smc
Tam\'{a}s F\"{u}l\"{o}p\footnote{${}^\dagger$}
{\eightpoint email:\quad fulopt@poe.elte.hu}
}

\vskip 3pt
{
\baselineskip=13pt
\centerline{\it Institute for Theoretical Physics}
\centerline{\it Roland E\"{o}tv\"{o}s University}
\centerline{\it H-1117 Budapest, P\'{a}zm\'{a}ny
P. s\'{e}t\'{a}ny 1/A, Hungary}
}

\vskip 3pt
\centerline{\rm and}
\vskip 3pt

\centerline{\smc
Taksu Cheon\footnote{${}^\ddagger$}
{\eightpoint email:\quad cheon@mech.kochi-tech.ac.jp,
http://www.mech.kochi-tech.a.jp/cheon/} 
}

\vskip 3pt
{
\baselineskip=13pt
\centerline{\it Laboratory of Physics}
\centerline{\it Kochi University of Technology}
\centerline{\it Tosa Yamada, Kochi 782-8502, Japan}
}

\vskip 15pt

\abstract{%
{\bf Abstract.}\quad
The Schr{\"o}dinger operator with point interaction
in one dimension has a $U(2)$ family of self-adjoint
extensions.  We study
the spectrum of the operator and show that
(i) the spectrum is uniquely determined by the eigenvalues
of the matrix $U \in U(2)$ that characterizes the extension,
and that 
(ii) the space of distinct
spectra is given by 
the orbifold $T^2/\Z_2$ which is 
a M{\"o}bius strip with boundary.  
We employ a parametrization of $U(2)$ that admits
a direct physical interpretation 
and furnishes a coherent
framework to realize
the spectral duality and anholonomy recently
found.  This allows us to find that (iii)
physically  distinct point interactions
form a three-parameter quotient space of the $U(2)$ family.
}

\vskip 10pt
%
%
%
%
}
\bigskip
\ve


\pageheight{23cm}
\pagewidth{15.7cm}
\hcorrection{-1mm}
\magnification= \magstep1
\def\bsk{%
\baselineskip= 15pt plus 1pt minus 1pt}
\parskip=5pt plus 1pt minus 1pt
\tolerance 8000
\bsk


\bigskip
\noindent{\bf 1. Introduction}
\medskip

Quantum mechanical motion of a particle
subject to a point interaction on a line $\R$
is described by
the free Schr{\"o}dinger
(the Laplacian) operator,
    $$
    H =
    -{{\hbar^2}\over{2m}}{{\d^2}\over{\d x^2}} \ ,
    \eqn\ham
    $$
with one point perturbation.
This is
implemented by deleting a point, say $x = 0$ on the line,
and thereby considering
the family $\Omega$ of self-adjoint operators $H$ defined 
on proper domains in the Hilbert space
${\cal H} = L^2(\R\setminus\{0\})$.
The theory of self-adjoint extensions then
dictates
that the family $\Omega$ is given by
the group $U(2)$, which covers all 
allowable distinct point
interactions [\Seba].   Studies
show that the spectrum of the operator $H$ consists of
the essential spectrum
$[0, \infty)$ together with a
discrete spectrum having at most two
levels of bound states [\ABD]
(see also [\AGHH,\AK] and references therein).
Symmetries such as parity or time-reversal are used to classify the
family $\Omega \simeq U(2)$ in terms of their invariant subfamilies
[\ADK].

Recently, we 
have examined the spectral properties of
this simple system
and found a number of
interesting features
which are  usually ascribed to more 
complex systems.
These features include
duality in the spectra under
strong vs weak coupling exchange [\CSb,\TFC], anholonomy 
both in the phase
of states (the Berry
phase) and in levels under a cycle in $\Omega$
[\CHa,\EG], and the double degeneracy 
which leads to supersymmetry
[\CFT].  Meanwhile, a similar study has been made on a circle $S^1$
with point interaction [\FT],
where it is shown that
the spectrum of $H$ does not depend on the entire $U(2)$
parameters 
as one na{\"\i}vely expects.

The aim of the present
paper is to furnish
a comprehensive
picture of the spectral
structure of the entire family of the Schr{\"o}dinger
operators $H$ on a line $\R$
as well as on an interval $[l, -l]$ (under some innocuous boundary
conditions) with the point
$x = 0$ removed. 
Our main results are given in three theorems. Theorem 1 states
that the spectrum is uniquely determined by the eigenvalues of the
$U(2)$ matrix which characterizes the point interaction, and
Theorem 2 shows that, for the case of the interval,
the space
$\Sigma$ consisting of all distinct spectra is given by a M{\"o}bius
strip with boundary, while for the case of the line 
$\Sigma$ 
is a subspace
of it.  The key
observation to reach these statements is that the
set of $su(2)$ parity transformations on the operator $H$ which
preserve the spectrum [\TFC,\CFT] can be generalized in order
to narrow
the dependence from $U(2)$ down
to its subspace.  We also provide a
generalization in symmetry transformations in order to associate
a pertinent invariant subfamily to any point interaction in $\Omega$.
In our treatment emerges a 
natural parametrization of $\Omega$
which admits a direct
physical interpretation and furnishes a framework
to describe the above mentioned
features in a coherent manner.
As part of the physical interpretation given as Theorem 3,
we find a one-parameter gauge equivalence within $\Omega$
and conclude that
physically distinct point interactions form  
a three-parameter
quotient space of $\Omega$.

\bigskip
\noindent{\bf 2. Spectral structure}
\medskip

Let us first recall the description of the $U(2)$
family of self-adjoint operators $H$ [\TFC] (see also [\AG]).
The domain of such a self-adjoint operator $H$ is a subspace
of ${\cal H}$ specified by a boundary condition
at the missing point $x = 0$ on the line.
Let $\varphi$ be a state 
in the domain, and
consider the two-component boundary vectors
    $$
    \Phi :=
      \left( {\matrix{{\varphi (0_+)}\cr
                      {\varphi (0_-)}\cr}
             }
      \right),
    \qquad
    \Phi' :=
      \left( {\matrix{{ \varphi' (0_+)}\cr
                      {-\varphi' (0_-)}\cr}
             }
      \right) ,
    \eqn\vectors
    $$
where $0_+$ and $0_-$ denote the limits at $x = 0$ from the right
and the left, respectively.\note{$\varphi$ and
its derivative,
$\varphi'$, are required to be absolutely continuous on
$ \R \setminus \{ 0 \} $ (see [\AG]).}
In terms of a matrix $U \in U(2)$ the boundary condition
is then given as
    $$
    (U-I)\Phi+iL_0(U+I)\Phi'=0\ ,
    \eqn\unitrel
    $$
with some constant $L_0 \ne 0$ of length dimension, where $I$
denotes the unit matrix in $U(2)$.
We note that the self-adjointness
of $H$ is equivalent to the requirement of (global) probability
conservation, and that the constant $L_0$ adds no extra
freedom
to that given by $U$ [\CFT].
To indicate the $U(2)\bitt$-dependence of the operator $H$, we use the
notation $H_U$.

We now begin our discussion of the spectral structure of the
family $\Omega$ of the operators $H_U$ by providing the following
\smallskip
\noindent
{\bf Definition 1.}
A unitary transformation ${\cal X}:  {\cal H} \rightarrow {\cal H}$ is
called {\it a generalized symmetry} of the family $\Omega$ if, for
any $U \in U(2)$,
    $$
    {\cal X}^{-1} H_U {\cal X} = H_{U_{\cal X}}\ ,
    \eqn\aaa
    $$
for some $ U_{\cal X} \in U(2) $.

\smallskip
\noindent
We note that
condition (\aaa) embodies two requirements: first,
the domain of $H_U$ is mapped into the domain of $H_{U_{\cal X}}$,
and secondly, ${\cal X}^{-1} H_U {\cal X}$ acts on this new domain as
the differential operator (\ham).  Note also that 
the two operators $H_U$ and $H_{U_{\cal X}}$ share the same spectrum.

\smallskip

The following lemmas will be useful in proving Theorem 1.

\smallskip
\noindent
{\bf Lemma 1.}
{\it The operators ${\cal P}_j$ $ ( j = 1, 2, 3 ) $ defined as}
    $$
    \eqalign{
    & ({\cal P}_1\varphi)(x) := \varphi( - x)\ , \cr
    & ({\cal P}_2\varphi)(x) :=
     i[\Theta(-x) - \Theta(x)]\varphi(-x)\ , \cr
    & ({\cal P}_3\varphi)(x) := [\Theta(x) -
    \Theta(-x)]\varphi(x)\ ,
    }
    \eqn\rtrsf
    $$
{\it (where $\Theta$ denotes the Heaviside step function) are generalized
symmetries.  Further, they are 
parity-type operators (i.e., ${\cal P}_j^2 = \qidH$, $\qP_j \neq \pm
\qidH$) 
and satisfy the $su(2)$ commutation relations $[{\cal P}_j,
{\cal P}_k] = 2i\sum_{l = 1}^3 \epsilon_{jkl} {\cal P}_l$ and the
anticommutation relations $ \{ \qP_j, \qP_k \} = 2 \delta_{jk} \qidH$.}
\smallskip

\noindent{\it Proof.}
It is straightforward to check that these operators 
are unitary 
and parity-type, fulfilling the stated commutation and anticommutation
relations. To show that
they are 
generalized symmetries, let us observe that, under a
${\cal P}_j$, the boundary vectors (\vectors) change as $\Phi \mapsto
\sigma_j
\Phi$ and $\Phi' \mapsto \sigma_j \Phi'$, where $\sigma_j$s denote the Pauli
matrices. In the boundary condition
(\unitrel), this change can be absorbed by the change in the matrix $U$ as
    $$
    U \longmapsto U_{{\cal P}_j} := \sigma_j\, U \, \sigma_j\ .
    \eqn\uchange
    $$
This implies that a ${\cal P}_j$ maps the domain of an $H_U$ to the domain
of $H_{U_{{\cal P}_j}}$ with $U_{{\cal P}_j}$ given in (\uchange)
(clearly, ${\cal P}_j$s preserve the smoothness properties mentioned in
Footnote 1, too). It is also easy to see that $ \qP_j H_U^{} \qP_j $
remains the differential operator (\ham) on this new domain, since,
under any of the transformations (\rtrsf), $\varphi$ acquires merely
an overall complex phase factor that is constant on both $\R_+$ and
$\R_-$. {\it Q.E.D.}

\smallskip

The three transformations
defined
above are not the only
parity-type generalized symmetries. Indeed,
operators given by the linear combinations of the three,
    $$
    \qP := \sum_{j = 1}^3 c_j \, \qP_j \hskip 8.5ex \hbox{with}
    \quad c_j \in \R \, , \hskip 3.5ex \sum_{j = 1}^3 c_j^2 = 1 \ ,
    \eqn\generalpar
    $$
are all generalized symmetries and fulfill the parity property $\qP^2 =
\qidH$, where now the induced transformation on $U$ reads
    $$
    U  \longmapsto U_{\cal P} = \sigma \bitt U \bitt \sigma\ , \qquad
    \sigma := \sum_{j = 1}^3 c_j\, \sigma_j\ .
    \eqn\gensigm
    $$
We therefore arrive at

\smallskip
\noindent
{\bf Lemma 2.}
{\it {}For any $su(2)$ element $\sigma$ normalized as $\sigma^2 = I$,
and for any $ U \in U(2) $,
$ H_U $ and $ H_{\sigma \bit U \bit \sigma} $ share an identical
spectrum.}
\smallskip

Using Lemma 2, we now show

\smallskip
\noindent
{\bf Theorem 1.}
{\it The spectrum of the Schr{\"o}dinger operator $H_U$ is uniquely
determined by the eigenvalues of the matrix $U$.}
\smallskip

\noindent
{\it Proof.}
Let $e^{i\theta_+}$ and $e^{i\theta_-}$ with $\theta_\pm \in [0, 2\pi)$
be the two eigenvalues of the unitary matrix $U$.  These eigenvalues arise
in the 
matrix,
    $$
    D = \pmatrix{e^{i\theta_+} & 0 \cr 0 & e^{i\theta_-} \cr}\ ,
    \eqn\diagmat
    $$
which appears when one diagonalizes
    $$
    U = V^{-1} D V\ ,
    \eqn\decu
    $$
with an appropriate $V \in SU(2)$. To proceed, let us set
    $$
    D = e^{i\xi}\, e^{i\rho\sigma_3}\ , \qquad
    \xi = {{\theta_+ + \theta_-}\over 2}, \qquad
    \rho = {{\theta_+ - \theta_-}\over 2}\ ,
    \eqn\dparm
    $$
to rewrite (\decu) as
    $$
    U = e^{i\xi}\,e^{i\rho\, V^{-1}\sigma_3 V} \ .
    \eqn\ndecu
    $$
Note that $V^{-1}\sigma_3 V$ in the exponent is just an
element of $su(2)$ obtained by the rotation of $\sigma_3$ with respect to
an axis determined by $V$.  Note also that, since
$\sigma = {1\over i} e^{{\pi\over 2}i\sigma} = \sigma^{-1}$,
the product $\sigma \sigma_3 \bit \sigma$ is an
element of $su(2)$ obtained by the rotation of $\sigma_3$ with
respect to $\sigma$ by the angle $\pi$.  This implies that,
to a given $V$, one can always find some $\sigma$ such that
$V^{-1}\sigma_3 V = \sigma \sigma_3 \bit \sigma$ holds.
With such $\sigma$ we now have
    $$
    U = e^{i\xi}\,e^{i\rho\, \sigma \sigma_3 \sigma}
    =  \sigma\, D\, \sigma \ .
    \eqn\nndecu
    $$
Lemma 2 then ensures that the spectrum of $H_U$ coincides with the
spectrum of $H_D$. {\it Q.E.D.}
\smallskip

{}From this theorem we obtain
\smallskip
\noindent
{\bf Corollary 1.}
{\it
A point interaction
characterized by $U$ possesses the
{\rm isospectral subfamily}
    $$
    \Omega(D) := \left\{ \bit H_{ V^{-1} D V } \, \vert \,
     V \in SU(2) \bit \right\}\ ,
    \eqn\noaaa
    $$
where $D$ is the diagonal eigenvalue matrix in the decomposition {\rm
(\decu)} of $U$.
The isospectral subspace $\Omega(D)$
is homeomorphic to the coadjoint orbit of
$SU(2)$ passing through the element $e^{i\rho\sigma_3}$, and
hence $\Omega(D) \simeq S^2$ except for the case
$D = e^{i\theta }\cdot I$ $(\theta \in [0, 2\pi))$
for which $\Omega(D)$
consists of
$D$ alone.}
\smallskip


\noindent
We mention that the exceptional cases
($\theta = \theta_+ = \theta_-$)
occur at 
    $$
    U = e^{i\theta }\cdot I
    \ ,
    \qquad \theta \in [0, 2\pi)\ ,
    \eqn\sdsubfamily
    $$
which form what we call the {\it self-dual subfamily}
$\Omega_{\rm
SD} \simeq U(1)$ in the entire set of point interactions $\Omega \simeq
U(2)$ (see also Proposition 3 and the remark which follows). 

Clearly, the two eigenvalues of $U$
appearing in $D$ are interchangeable, and this is realized for
$\Omega(D)$ by setting, {\it e.g.,} $V \mapsto i\sigma_2 V$.
Thus, if we write
$D = D(\theta_+, \theta_-)$ for the diagonal matrix $D$ in (\diagmat),
we have

\smallskip
\noindent
{\bf Corollary 2.}
{\it
The two isospectral subfamilies associated with $D(\theta_+, \theta_-)$
and $D(\theta_-, \theta_+)$ are identical,
    $$
    \Omega(D(\theta_+, \theta_-)) = \Omega(D(\theta_-, \theta_+))\ ,
    \eqn\noaab
    $$
and hence the spectrum occurring at
$D(\theta_+, \theta_-)$ and that occurring
at $D(\theta_-, \theta_+)$ are the same.
}
\smallskip

The spectral feature discussed above is seen 
in the discrete spectrum, but it is
largely obscured because 
the spectrum consists
mostly of the continuous spectrum
$[0, \infty)$.   However, the structure
becomes manifest if one considers,
instead of a line, a box (interval) 
on which the entire
spectrum becomes discrete.  This can be done by
imposing a boundary condition at both 
ends of the box
in such a way
that it does not
affect the consequences of the 
operations of ${\cal P}$ in (\generalpar).  
Specifically,
if we let the interval $[-l, l]$ be the box
where the point interaction
is placed at $x = 0$, then we seek for boundary
conditions at $x = \pm l$ which remain invariant under any of the
transformations induced by ${\cal P}$.  These are given by

\smallskip
\noindent
{\bf Proposition 1.}
{\it The boundary conditions at $x = \pm l$ which are left
unchanged under any of the
transformations induced by ${\cal P}$ (and hence provide
a domain for $H$ so that the entire discrete spectrum exhibits
the spectral structure manifestly) are
    $$
    \varphi (l) + L\, \varphi' (l) = 0\ , \qquad
    \varphi (- l) - L\, \varphi' (- l) = 0\ ,
    \eqn\goodbd
    $$
where $L \in (-\infty, \infty) \cup \{\infty\}$ is an arbitrary
parameter.
}
\smallskip
\noindent
{\it Proof.}
The operator $H$ remains self-adjoint if the boundary
condition at $x = \pm l$ ensures the probability conservation,
and this is exactly the demand we used to obtain
the boundary condition (\unitrel) at
$x = 0$.
(More precisely, one needs to require further that
the probability current vanish at the both ends, but this will be seen
to be satisfied at the end.)
This suggests
that, if we use the boundary vectors similar to (\vectors),
    $$
    \Psi :=
      \left( {\matrix{{\varphi (l)}\cr
                      {\varphi (-l)}\cr}
             }
      \right),
    \qquad
    \Psi' :=
      \left( {\matrix{{ \varphi' (l)}\cr
                      {-\varphi' (-l)}\cr}
             }
      \right) ,
    \eqn\bvectors
    $$
the
boundary conditions at the ends
can be given analogously as
    $$
    (\tilde U-I)\Psi+iL_0( \tilde U+I)\Psi'=0\ ,
    \eqn\bunitrel
    $$
in terms of a matrix $\tilde U \in U(2)$ characterizing
the two ends.  The transformation 
of the operator ${\cal P}$ on the
boundary vectors (\bvectors) is the same as before, and hence
it induces the same action
$\tilde U  \mapsto \tilde
U_{\cal P} = \sigma\, \tilde U\, \sigma$ on the matrix $\tilde U$.
Thus, the required boundary condition must satisfy
$\sigma\, \tilde U\, \sigma = \tilde U$, that is, we find
$\tilde U = e^{i\theta}\cdot I$ for $\theta \in [0, 2\pi)$.
Putting
$L = L_0 \cot{{\theta}\over 2}$ we obtain the statement.
{\it Q.E.D.}
\smallskip

\noindent
We remark that both
the Dirichlet condition
$\varphi(l) =\varphi(-l) = 0$ and
the Neumann
condition
$\varphi'(l) =\varphi'(-l) = 0$ are of the type (\goodbd).

If we now introduce the space of distinct spectra,
    $ \bittt
    \Sigma := \{ \bittt {\rm Spec\bitt} (H_U) \bittt \vert \bittt
    U \in U(2) \bit \}\bit , \bittt
    $
then from the foregoing argument we find that
$\Sigma$ is a subspace of 
the torus $T^2 = S^1 \times S^1 = \{ (\theta_+, \theta_-)\}$
subject to the identification 
$(\theta_+, \theta_-) \equiv (\theta_-, \theta_+)$.
The quotient space obtained by the identification 
is the orbifold
$T^2/\Z_2$ which is the domain of the triangle shown in Fig.1. 
The elementary observation in Fig.1 leads to

\smallskip
\noindent
{\bf Theorem 2.}
{\it  The spectral space $\Sigma$ of point interactions 
is a subspace of the orbifold $T^2/\Z_2$ which is homeomorphic to
a M{\"o}bius strip with boundary.  In particular, for the 
box $[l, -l]$ the spectral  
space $\Sigma$ is the entire $T^2/\Z_2$.
}
\smallskip
\noindent
{\it Proof.}
The first half is already shown (see Fig.1).  
To show the second half, we
observe that for an isospectral subfamily 
$\Omega(D)$ the spectrum is determined by
the boundary condition (\unitrel), which splits into
    $$
    \varphi (0_+) + L_+\, \varphi' (0_+) = 0\ , \qquad
    \varphi (0_-) - L_-\, \varphi' (0_-) = 0\ ,
    \eqn\spltbd
    $$
where we have used
    $$
    L_\pm := L_0 \cot{{\theta_\pm}\over 2}\ .
    \eqn\conbi
    $$
Then for the 
box $[l, -l]$ the problem boils down 
to determining the spectrum of the operator
in two separate boxes, $[-l, 0_-)$ and
$(0_+, l]$, under the combined boundary conditions,
(\goodbd) and (\spltbd).  
{}For the interval $(0_+, l]$, for instance, the positive
spectrum $E = \hbar^2 k^2/(2 m)$ 
is determined by the condition,
$\tan kl = k(L - L_+)/(1 + k^2 L L_+)$, which admits
a distinct set of solutions for different 
$L_+$ under fixed $L$.  It thus follows that to each pair
$(L_+, L_-)$ or $(\theta_+, \theta_-)$ modulo the
exchange $\theta_+ \leftrightarrow \theta_-$ there arises
a distinct spectrum. {\it Q.E.D.}
\smallskip

\pageinsert
\vskip 1cm
\let\picnaturalsize=N
\def\picsize{5cm}
\def\picfilename1{f1.epsf}
\input epsf
\ifx\picnaturalsize N\epsfxsize \picsize\fi
\hskip 3.7cm\epsfbox{\picfilename1}
\vskip 0.2cm
\abstract{%
{\bf Figure 1.}
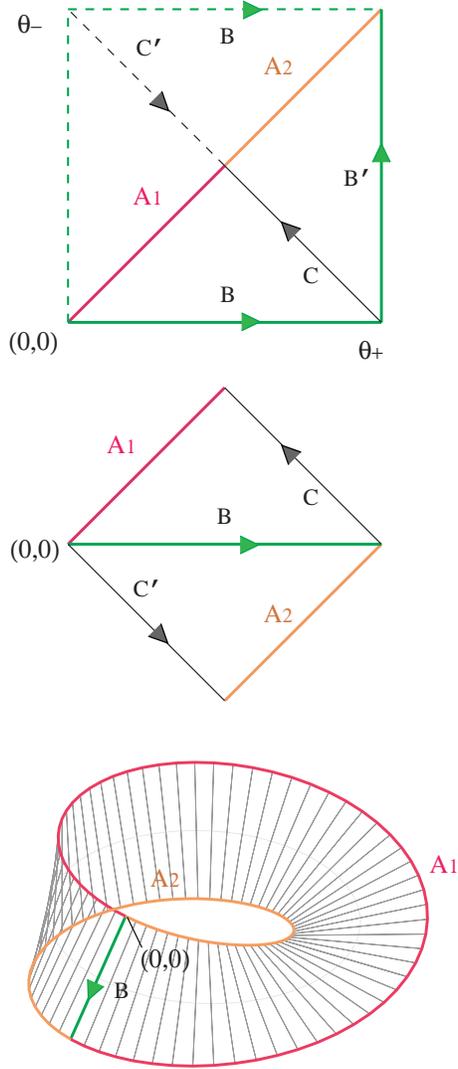
In the top figure, the spectral space $\Sigma$
is the triangle surrounded
by edges $A_1 + A_2$, $B$ and $B'$.
We divide this triangle into two
subtriangles $B$--$C$--$A_1$ and $B'$--$C$--$A_2$.
Since the latter subtriangle
is spectrally identical to its dual
image $B$--$C'$--$A_2$, $\Sigma$ can be represented
by the square $A_1$--$C'$--$A_2$--$C$ in the middle figure.
When the two spectrally identical edges
$C$ and $C'$ are stitched together with the right orientation,
we obtain
the M{\"o}bius strip with boundary $A_1$--$A_2$
representing the self-dual subfamily $\Omega_{\rm SD}$
(the bottom figure).
}
\vskip 0.5cm
\endinsert

We have seen that the product form (\decu) for
the matrix $U$
furnishes a useful parametrization
for the point interaction in one dimension,
where the spectral property resides solely in the diagonal
part $D$.  The adjoint part $V$, on the other hand, may be used
to provide a parity transformation pertinent to the point interaction
as follows.

\smallskip
\noindent
{\bf Proposition 2.}
{\it  To a point interaction specified by $U$ there is a parity
operator ${\cal P}$ of the form {\rm (\generalpar)}
whose action leaves $U$ invariant.  The operator ${\cal P}$
is unique (up to the sign) except when $U \in \Omega_{\rm SD}$ for 
which ${\cal P}$ is arbitrary.
}
\smallskip
\noindent
{\it Proof.}
Consider the $su(2)$ element $\sigma$ in (\gensigm) given by
    $$
    \sigma = \sigma(V) := V^{-1}\sigma_3 V\ ,
    \eqn\sfbyu
    $$
where $V$ is the $SU(2)$ matrix appearing in (\decu)
for the the diagonalization of
the matrix $U \not\in \Omega_{\rm SD}$.
Note that in (\decu) the matrix $V$ is determined
only up to the left action $e^{i\chi\sigma_3} V$, 
but this ambiguity
does not affect in specifying $\sigma$ in (\sfbyu).
We now expand $\sigma(V)$ in the $su(2)$ basis as
$\sigma(V) = \sum_{j = 1}^3 c_j(V)\, \sigma_j$ and define
the corresponding parity operator,
$$
{\cal P}(V) := \sum_{j = 1}^3 c_j(V)\, {\cal P}_j\ .
\eqn\sppar
$$
We then see at once
that, under the transformation induced
by ${\cal P}(V)$, the matrix $U$ is left invariant,
$\bittt \sigma(V)\, U\,\sigma(V)\ = U$.  The parity
$\bitt -{\cal P}(V) \bitt$ corresponding to $\bitt -\sigma(V)\bitt$ also
leaves $U$ invariant.
{}For $U \in \Omega_{\rm SD}$, it is obvious that
any $\sigma$, and hence
any ${\cal P}$
in (\generalpar) leaves $U$ invariant. {\it Q.E.D.}
\smallskip

The content of Proposition 2 may equally be stated as

\smallskip
\noindent
{\bf Proposition 2'.}
{\it The Schr{\"o}dinger
operator $H_U$ commutes with
a parity operator ${\cal P}$ given by {\rm (\generalpar)},
$ \bittt \left[\, H_U, {\cal P}\,\right] = 0\, , \bittt $
where for
$U \not\in \Omega_{\rm SD}$
the operator ${\cal P}$ is uniquely determined as
${\cal P} = {\cal P}(V)$ (up to sign) in {\rm (\sppar)},
while for $U \in \Omega_{\rm SD}$
it is arbitrary.
}
\smallskip

\noindent
We note that, for an $H_U$ and the parity operator 
$\qP$ commuting with it, 
the Hilbert space $\qH$ can be decomposed into
two orthogonal closed linear subspaces, $ \qH = \qH_+ \oplus \qH_- $,
where $\qH_+$ and $\qH_-$ are the eigenspaces
of $\qP$ corresponding to the eigenvalues $1$ and $-1$, respectively.
The nondegenerate eigenfunctions of $H_U$ belong to either $\qH_+$ or
$\qH_-$.  {}For doubly degenerate eigenvalues, the eigenfunctions can be
chosen such that one belongs to $\qH_+$ and the other to $\qH_-$.
Note that, since
the eigenvalue equation is a second order differential equation on both
half lines,
the eigenvalues of $H_U$ are at most doubly degenerate. 
Namely, these degenerate solutions
contain two free constants each, and 
the boundary condition
(\unitrel) reduces this four-parameter freedom to a two-parameter one.
These statements are valid for the non-normalizable eigenfunctions
(scattering states) of $H_U$, too, in the rigged Hilbert space sense
(note that the definition (\rtrsf), and correspondingly the definition of
$\qP(V)$, can be extended to any $ \bitt \R \setminus \{ 0 \} \to \C
\bitt $ function
in a natural way, 
which involves the natural extension of $\qH_+$ and
$\qH_-$).  

A distinguished family of generalized symmetries which interchange the
subspaces 
$\qH_+$ and $\qH_-$ exist, that is, 

\smallskip
\noindent
{\bf Proposition 3.}
{\it  {}For an $H_U$ and the associated 
parity operator $\qP$ commuting with
it, there exists a $U(1)$ family of 
generalized symmetries $\qD$ such
that each $\qD$ maps $\qH_+$ to $\qH_-$ and vice versa, and satisfies
$(U_\qD)^{}_\qD = U$.
}
\smallskip
\noindent
{\it Proof.}  
Consider the generalized symmetries $\qD$ corresponding to the
$U(1)$ family of $su(2)$ elements, 
    $$
    \sigma_{\qD} := 
    V^{-1}\,\tilde\sigma(\phi)\, V\ ,
    \eqn\onepfmly
    $$
where we have defined 
$\tilde\sigma(\phi) 
=  \cos \phi\, \sigma_1 + \sin \phi\, \sigma_2$ for 
$\phi \in [0, 2\pi)$,
and introduced $\qD := \sum_{j=1}^3 c_j' \qP_j$ using the expansion
$ \, \sigma_\qD = \sum_{j=1}^3 c_j' \sigma_j \, $ of $\sigma_\qD$.
On $U = V^{-1}\, D(\theta_+, \theta_-)\,V$ these $\qD$ induce 
     $$
     \eqalign{
     U \longmapsto  U_\qD 
     &= \sigma_{\qD}\, U\,\sigma_{\qD} \cr
     &= V^{-1}\,\tilde\sigma(\phi) \, D(\theta_+, \theta_-)\, 
     \tilde\sigma(\phi) \, V
     = V^{-1}\, D(\theta_-, \theta_+)\,  V\ ,
     }
     \eqn\dualtr
     $$
and hence implement the interchange $\theta_+
\leftrightarrow \theta_-$. {}From this $(U_\qD)^{}_\qD = U$ is clear.
To prove that a $\qD$ maps any eigenfunction of $\qP$ to another
one with opposite eigenvalue, we show that $\{ \qP, \qD \} = 0$. Indeed,
from $ \, \{ \qP_j, \qP_k \} = {\rm Tr} \bittt ( \sigma_j \sigma_k )
\bitt \qidH \, $ and (\sppar) it follows that
    $$
    \textstyle
    \{ \qP, \qD \} = {\rm Tr} \bittt \left( \sum_{j=1}^3 c_j \sigma_j
    \bittt \sum_{k=1}^3 c_k' \sigma_k^{} \right) \bit \qidH =
    {\rm Tr} \bitt \left( \sigma \sigma_\qD \right) \bit \qidH
    = {\rm Tr} 
    \bittt ( \sigma_3 \bit \tilde{\sigma}(\phi)  ) \bitt \qidH
    = 0\, .
    \eqn\no
    $$
{\it Q.E.D.}
\smallskip
  
\noindent
Hence, in the light of these properties, $\qD$ may be called {\it
duality transformation}.  The duality found in [\TFC,\CFT] is a special
case of $\qD$.

The role of the point interaction and the parity operator
in Proposition 2 can be reversed to obtain

\smallskip
\noindent
{\bf Proposition 4.}
{\it
To a parity operator ${\cal P}$ given in {\rm (\generalpar)}
there is a subfamily
of point interactions which are left
invariant under ${\cal P}$.
{}For any ${\cal P}$ the subfamily $\Omega_{\cal P}$
is homeomorphic
to a torus $T^2$.
}
\smallskip
\noindent
{\it Proof.}
The subfamily $\Omega_{\cal P}$ is given by
    $$
    \Omega_{\cal P} := \left\{\, U \in U(2) \ \vert \bitttt
    \sigma\bittt U\,\sigma = U \, \right\}\ ,
    \eqn\noaad
    $$
where $\sigma$ is determined from ${\cal P}$ by (\gensigm).
The matrices $U$ belonging to $\Omega_{\cal P}$ are then found to be
of the form,
    $$
    U = e^{i\xi}\,e^{i\rho \sigma}\ ,
    \qquad \xi \in [0, \pi)\ , \quad
    \rho \in [0, 2\pi)\ ,
    \eqn\noaae
    $$
which is homeomorphic to a torus $T^2$
for any ${\cal P}$.  {\it Q.E.D.}%
\note{
To derive the above results one may consider, 
instead of the parities
${\cal P}$ in (\generalpar), 
more general transformations $\qF_W$ given by
$ (\qF_W
\psi) (x) := W_{11} \Theta(x) \psi(x) + W_{12} \Theta(x) \psi(-x) + W_{21}
\Theta(-x) \psi(-x) + W_{22} \Theta(-x) \psi(x) 
$ with the matrix $W$ of the coefficients $W_{ij}$ belonging to 
$U(2)$.  These generalized symmetries $\qF_W$ 
realize the arbitrary
boundary conjugations
$ U_\qX = W U W^{-1}$ and obey several useful
properties, such as 
$ \ \qF_{W_1 W_2} = \qF_{W_1} \qF_{W_2} \, $ and $\ \qF_{
\lambda_1 W_1 + \lambda_2 W_2 } 
= \lambda_1 \qF_{W_1} + \lambda_2 \qF_{W_2}$ for
$\lambda_1, \lambda_2 \in \C$.
}
\smallskip

\noindent
{}For instance, if we choose ${\cal P} = {\cal P}_1$,
the subfamily $\Omega_{{\cal P}_1}$
is just the set of parity invariant (left-right
symmetric) point interactions in the usual sense of the word.
If, on the other hand,
we choose ${\cal P} = {\cal P}_3$, then the resultant subfamily
$\Omega_{{\cal P}_3}$ becomes the so-called separated
subfamily where no probability
flow through the gap $x = 0$ is allowed.
One may also choose for ${\cal P}$ the one ${\cal P}(V)$
that corresponds to
a specific $U$.
The invariant subfamily $\Omega_{{\cal P}(V)}$ then contains
$U$ by construction, and becomes a subfamily
pertinent to the point interaction characterized by $U$.
One then finds from Propositions 3 and 4 that
$\Omega_{{\cal P}(V)} \simeq T^2$ except when
$U \in \Omega_{\rm SD}$ for which $\Omega_{{\cal P}(V)}$ coincides
with the entire family $\Omega \simeq U(2)$.

The self-dual subfamily $\Omega_{\rm SD}$ has also
the following distinguished characteristics:

\smallskip
\noindent
{\bf Proposition 5.}
{\it
{}For any point interaction belonging to $\Omega_{\rm SD}$ 
(\ie $ U \in
\Omega_{\rm SD}$), all eigenvalues of $H_U$ (including the
generalized ones) are doubly degenerate.
}
\smallskip
\noindent
{\it Proof.}
{}For any $ U \in \Omega_{\rm SD}$, 
we have from Proposition 2' that 
$
\bitt [\, H_U, {\cal P}_j\,] = 0 \bitt 
$ 
for $j = 1$, 2, 3.  This implies
that, on any eigenspace of $H_U$, a representation of $su(2)$ formed by
$\{ {\cal P}_j\}^{}_{j = 1, 2, 3} $ is given. Since an eigenfunction of,
say,
${\cal P}_1$ cannot be an eigenfunction of ${\cal P}_2$, the eigenspaces
of $H_U$ must be doubly degenerate. 
This argument is valid for the
generalized eigenvalues (scattering state energies) and the corresponding
eigenspaces as well.
{\it Q.E.D.}
\smallskip
\noindent
The double degeneracy implies that the system
with point interaction
belonging to $\Omega_{\rm SD}$ may be regarded as
supersymmetric.  As
shown in [\CFT], this is in fact the case for $U = -I$,
where the energy of the two bound states
vanishes yielding an $N = 2$ Witten model with a
\lq good SUSY\rq{} [\Junker].  Generically,
however, the ground state energy of the system is nonvanishing
and the system is not supersymmetric even though
it admits a formally
supersymmetric reformulation for any $U$ of $\Omega_{\rm SD}$.
The obstacle for being supersymmetric
is the fact that the presumed supercharges
are not self-adjoint unless $U = -I$.

We have learned that the spectrum of the
operator $H_U$ 
is determined by the two parameters
in $D$ in the decomposition $U = V^{-1} D V$, and that, in particular,
for the box the space $\Sigma$ of the spectra is given by a M{\"o}bius
strip.  
One can proceed further and assign more general
physical meaning to the parameters in the matrix
$U$.  To see this we first rewrite the boundary condition
(\unitrel) using the decomposition
as
    $$
    V \Phi +   \pmatrix{L_+ & 0 \cr 0 & L_- \cr} V
    \Phi' = 0\ ,
    \eqn\newunitrel
    $$
with $L_\pm$ given in (\conbi).  
We further parametrize $V$ by the Euler angles
(with the first factor $e^{i\chi\sigma_3}$ 
which does not affect $U$
being dropped),
    $$
    V = e^{i{\mu\over 2}\sigma_2} e^{i{\nu\over 2}\sigma_3} \ ,
    \qquad \mu \in [0, \pi], \quad \nu \in
    [0, 2\pi) \, ,
    \eqn\contv
    $$
and thereby present
    $$
    \left\{(L_+, \, L_-, \, \mu, \, \nu)\, \big\vert\,
    L_\pm \in (-\infty, \infty) \cup \{\infty\},
    \,  \mu \in
    [0, \pi], \, \nu \in [0, 2\pi) \right\}\ ,
    \eqn\ppset
    $$
as a basic set for the parametrization of the point
interactions on a line.
On account of the double specification of
the eigenvalues of
$U$, the set (\ppset)
is in a two-to-one
correspondence to
$U$, providing a double covering%
\note{
This redundancy is introduced to 
avoide unwanted discontinuity which arises when we
study the response, such as the level change or
the anholonomy phase,
of the system under smooth changes
over $\Omega$.
}
of the whole family
$\Omega \simeq U(2)$ (see Fig.2).
We then have

\smallskip
\noindent
{\bf Theorem 3.}
{\it
The parameters in the set {\rm (\ppset)} possess the following physical
properties:

\item{{\rm (i)}}
The two parameters $L_\pm$ furnish two independent
length scales to the point interaction.

\item{{\rm (ii)}}
The angle
$\nu$ is physically irrelevant (unobservable).

\item{{\rm (iii)}}
The angle
$\mu$
measures the extent of mixture of states between the
positive and negative half lines.

}
\smallskip
\noindent
{\it Proof.}
(i) is evident because in the boundary condition (\newunitrel)
$L_\pm$ are the only
parameters with length dimension.
To show (ii), we observe from (\contv) and (\newunitrel)
that the angle $\nu$
can be absorbed by introducing the new vectors
$e^{i\nu\sigma_3/2}\Phi$ and $e^{i\nu\sigma_3/2} \Phi'$
which arise if we replace
$\varphi (0_\pm) \mapsto e^{\pm i\nu/2}\varphi (0_\pm)$ and
$\varphi' (0_\pm) \mapsto e^{\pm i\nu/2}\varphi' (0_\pm)$.
This is implemented by
the $U(1)$ phase transformation (gauge transformation) on the state,
    $$
    \varphi(x) \longmapsto
    e^{{i\over\hbar}\vartheta(x)}\varphi(x)\ ,
    \qquad
    \vartheta(x)
    := {\nu\over 2}\, \left[\Theta(x) - \Theta(-x)\right]\hbar\ .
    \eqn\ugphs
    $$
Since the phase shift $\vartheta(x)$
is constant over $\R\setminus\{0\}$, and
since the phase gap (which occurs at the missing point
$x = 0$) cannot be
observed on a line,\note{
However, 
a phase gap may be observed, for instance, on a circle by
interference.
}
the transformed state is equivalent to the original state
in quantum theory, that is,
the angle $\nu$ is irrelevant physically.
Finally, (iii) is also evident in the boundary condition (\newunitrel)
because the factor $e^{i{\mu\over 2}\sigma_2}$ mixes the two rows
of the boundary vectors by rotation according to the angle $\mu$.
 {\it Q.E.D.}
\smallskip

\pageinsert
\vskip 1cm
\let\picnaturalsize=N
\def\picsize{6cm}
\def\picfilename1{f2.epsf}
\input epsf
\ifx\picnaturalsize N\epsfxsize \picsize\fi
\hskip 2.2cm\epsfbox{\picfilename1}
\vskip 0.2cm
\abstract{%
{\bf Figure 2.}
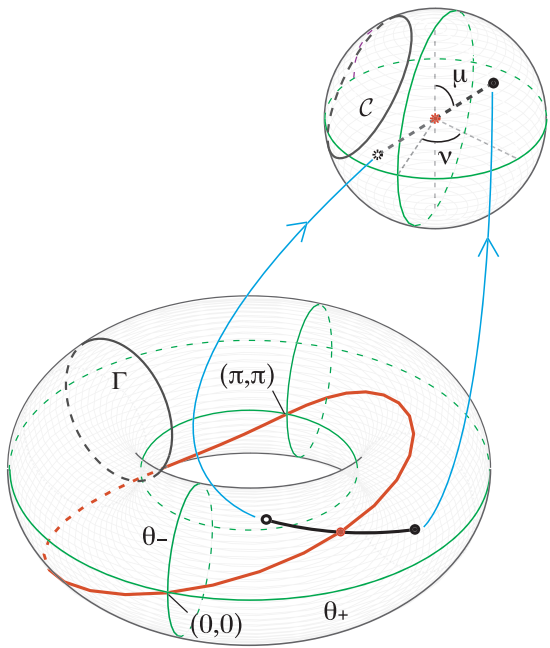
The parameter space
$\{( \theta_+, \theta_-, \mu, \nu)\}$ is
a product of the spectral torus $T^2$ 
specified by the angles
$(\theta_+, \theta_-)$
and the isospectral sphere $S^2$ 
specified by the angles
$(\mu, \nu)$ with
radius $\rho = (\theta_+ - \theta_-)/2$ (cf.~Corollary 1)
which collapses to a point for the
self-dual case $\theta_+ = \theta_-$.
A cyclic path ${\cal C}$ on the sphere yields a phase anholonomy
(the Berry phase)
proportional to the area enclosed by ${\cal C}$
due to the degeneracy
present at the center of the sphere.  A cyclic path $\Gamma$
on the torus, on the other hand, yields a level anholonomy
(level shifts) if $\Gamma$ is homotopically nontrivial.
A generic cycle is a combination of the two, and hence yields
an anholonomy in both phase and level.
The parametrization shown here provides a double covering
of the entire family $\Omega \simeq U(2)$, where
the two antipodal points on the spheres equidistant from
the self-dual line $\theta_+ = \theta_-$ are identified.  This
identification determines the spectral space $\Sigma$ to be
given by $T^2/{\Z}_2$ which is a M{\" o}bius strip with
boundary.
}
\vskip 0.5cm
\endinsert

\noindent
An important point to be noted here is that the existence of the
one-parameter gauge equivalence within
$\Omega$ implies that point interactions which are
distinct physically --- not just on the spectral basis --- 
form a three-parameter quotient space of
$\Omega$.

The properties stated
in Theorem 3 can be seen explicitly in
the  solutions of the Schr{\"o}dinger equation
under the operator $H_U$ being the Hamiltonian.  {}For instance,
the bound states allowed under $H_U$ on the line are given by
    $$
    \varphi_\kappa (x) = \ddef{ A_\kappa^- e^{\kappa x} , }{x < 0}{
    B_\kappa^+ e^{- \kappa x}, }{x > 0}
    \eqn\bounds
    $$
where $\kappa$ determines the bound state energy
$E_{\rm bound} = - \hbar^2\kappa^2/(2 m)$, and the constants
$A_\kappa^-$ and $B_\kappa^+$
are subject to the normalization condition
$\vert A_\kappa^- \vert^2 + \vert B_\kappa^+ \vert^2 = 2\kappa$.
A nonvanishing solution is then ensured if
   $$
   \kappa = {1\over L_+} \qquad \hbox{or} \qquad
   \kappa = {1\over L_-}\ ,
   \eqn\bsscale
   $$
which shows that there exist two bound states if $L_+ > 0$ and $L_-
> 0$, and one if $L_+ L_- < 0$, and none if
$L_+ < 0$ and $L_- < 0$.  The parameters $L_\pm$
thus give (in case they are positive)
the scales of the trapped particle.
In terms of (\contv) the coefficients are found to be
    $$
    \pmatrix{ B^+_\kappa \cr A^-_\kappa }
    = \sqrt{\frac{2}{L_+}}
    \pmatrix{ e^{ -2i  \nu  } \cos \frac{\mu}{2}
    \cr \sin \frac{\mu}{2} }\ ,
    \qquad
    \pmatrix{ B^+_\kappa \cr A^-_\kappa }
    = \sqrt{\frac{2}{L_-}} \pmatrix{
    - e^{ -2i\nu} \sin \frac{\mu}{2} \cr
    \cos \frac{\mu}{2} }\ ,
    \eqn\boundscoefb
    $$
for $\kappa = 1 / L_+$ and $1 / L_-$, respectively.
Note that the relative phase factor
$e^{ -2i\nu}$ attached to the coefficients
of the states on the positive half line can be removed by
(\ugphs).  Similarly,
the scattering states for the
particle (with
velocity
$v = \hbar k/m$) incident, say, from the positive side,
    $$
    \varphi_k\qr (x) = \frac{1}{\sqrt{2\pi}} \ddef{ t\qr_k e^{-ikx}
    \, , \quad \quad }{x < 0}{ e^{-ikx} + r\qr_k e^{ikx} \, , }{x > 0}
    \eqn\incr
    $$
have
the reflection and transmission
coefficients
$$
   \left( \matrix{r\qr_k\cr t\qr_k\cr} \right)
    = - {1\over{(1+i k L_+)(1+i k L_-)}}
    \left({\matrix{{1 + k^2 L_+ L_- - i k (L_+ - L_-) \cos\mu }\cr
                   {-i k (L_+ - L_-)\sin\mu\, e^{i\nu}} } }\right) \ .
    \eqn\scrform
$$
We observe that, in accordance with the interpretation,
the factor
$e^{i\nu}$ is simply the phase which is acquired by the
transmitted wave when
the incoming wave passes the point $x = 0$. 
We can also see that, unlike $\nu$, 
each of the other three parameters plays
an independent and physical role in the eigenstates of $H_U$.

Finally, let us illustrate the basic structure of
the $U(2)$ family by considering
a generic point interaction specified by $U$
in the $U(2)$
parameter space which is shown in Fig.2 as a product of 
a torus representing
$(\theta_+, \theta_-)$ and a sphere with
radius $\rho$ (see ($\nndecu$)
and Corollary 1)
representing
$(\mu, \nu)$.  
On this torus, two point interactions connected by 
the duality transformation (\dualtr) are represented by   
two equidistant points from the self-dual
loop, $\theta_+ = \theta_-$.
The double covering of the parametrization
implies that the two spheres attached to these dual
points are actually the same, with
antipodal points on the two spheres identified.
Under a cyclic process on the sphere one can expect a phase anholonomy
(the Berry phase) to arise,
since the spectrum
becomes degenerated at the center $\theta_+ = \theta_-$
which belongs to $\Omega_{\rm SD}$ (see Proposition 5).
One can also expect a level shift if the cycle is 
homotopically nontrivial 
on the torus
(see, {\it e.g.}, [\ER]).
The anholonomy both in phase and level has indeed been 
observed [\CFT] for
cycles  passing through $U = \sigma_3$,
that is, 
$(\theta_+, \theta_-) = (\pi, 0)$ and
$(\mu, \nu) = (0, 0)$.
We note that 
this point $U = \sigma_3$ is rather special because it
has the invariant parity
${\cal P}(V = I) =
{\cal P}_3$ and hence its
invariant subfamily is just the separated
subfamily $\Omega_{{\cal P}_3}$.  Further,  
its isospectral subfamily $\Omega(D = \sigma_3)$
is (the continuous part of) the
scale invariant subfamily $\Omega_{\rm W}$ [\ADK] in view of
the fact [\CFT] that such $U$ satisfies the condition for
scale invariance,
$\det (U \pm I) = \det (\sigma_3 \pm I) = 0$.  We stress,
however, that the anholonomy in phase and/or level is a 
generic phenomenon
observed for any cyclic process in the
parameter space $\Omega \simeq U(2)$.

\bigskip
\noindent
{\bf Acknowledgement:}
I.T. is indebted to S. Tanimura for useful comments.
T.C.~thanks members of the Theory Group of KEK for
the hospitality extended to him during his stay.
This work has been supported in part by
the Grant-in-Aid for Scientific Research (C)
(Nos.~10640301, 11640396 and 13640413) by
the Japanese
Ministry of
Education, Science, Sports and Culture.

\ve
\baselineskip= 15.5pt plus 1pt minus 1pt
\parskip=5pt plus 1pt minus 1pt
\tolerance 8000
\vfill\eject
\vfill\eject\immediate\closeout\reffile
\centerline{{\bf References}}\bigskip\frenchspacing%
\input refs.tmp\vfill\eject\nonfrenchspacing

\bye